\documentclass[11pt]{article}
\pdfoutput=1
\usepackage{jheppub}

\newcommand\nn{\nonumber}

\begin{document}

\preprint{MCTP-14-14}

\title{\boldmath On Exactly Marginal Deformations Dual to $B$-Field Moduli of IIB Theory on SE$_5$}

\author{Arash Arabi Ardehali,}
\author{Leopoldo A. Pando Zayas}

\affiliation{Michigan Center for Theoretical Physics, Randall Laboratory of Physics,\\
The University of Michigan, Ann Arbor, MI 48109--1040, USA}

\emailAdd{ardehali@umich.edu} \emailAdd{lpandoz@umich.edu}

\abstract{The complex dimension of the space of exactly marginal
deformations for quiver CFTs dual to IIB theory compactified on
$Y^{p,q}$ is known to be generically three. Simple general formulas
already exist for two of the exactly marginal directions in the
space of couplings, one of which corresponds to the sum of the
(inverse squared of) gauge couplings, and the other to the
$\beta$-deformation. Here we identify the third exactly marginal
direction, which is dual to the modulus
$\int_{\mbox{\footnotesize{2-cycle}}}B_{2}$ on the gravity side.
This identification leads to a relation between the field theory
gauge couplings and the vacuum expectation value of the gravity
modulus that we further support by a computation related to the
chiral anomaly induced by added fractional branes. We also present a
simple algorithm for finding similar exactly marginal directions in
any CFT described by brane tiling, and demonstrate it for the quiver
CFTs dual to IIB theory compactified on $L^{1,5,2}$ and the
Suspended Pinch Point.}

\maketitle \flushbottom

\section{Introduction}

\noindent The AdS/CFT correspondence has become one of the main
tools for understanding the strong coupling behavior of four
dimensional gauge theories. With minimal supersymmetry (eight
supercharges in AdS$_5$/CFT$_4$, or four supercharges when
conformality is broken) gauge theories appearing in the duality
exhibit rich enough behavior to be of great interest, yet are
sufficiently under computational control that testable non-trivial
information about them can be obtained from their gravitational
dual. One often arrives at such dual pairs by probing the singular
tip of a Calabi-Yau cone with a stack of D3-branes. The geometry
dual to the IR theory on the branes is of the form
AdS$_5\times$SE$_5$, with SE$_5$ the five-dimensional
Sasaki-Einstein base of the cone
\cite{Klebanov:1998,Acharya:1999,Morrison:1998}.

$Y^{p,q}$ are a countably infinite family of Sasaki-Einstein
5-manifolds that provide interesting supersymmetric
compactifications of IIB supergravity with simple quiver gauge
theories as CFT duals
\cite{Gauntlett:2004a,Gauntlett:2004b,Martelli:2004,Bertolini:2004xf,Benvenuti:2004}.
Since their discovery about a decade ago, these pairs of AdS/CFT
duals have been under considerable study and a variety of their
aspects, including the operators dual to classical strings
\cite{Benvenuti:2005} and supersymmetric branes on the gravity side
\cite{Canoura:2005uz}, is by now well understood. In particular, in
\cite{Herzog:2005,Benvenuti:2005a} it was realized that the
conformal manifold of the quiver gauge theories is (generically)
three-complex dimensional, and shortly afterwards all the three
gravity moduli dual to the exactly marginal directions were
identified \cite{Lunin:2005}.

However, on the field theory side, although two of the exactly
marginal directions were precisely known (one the
$\beta$-deformation and the other the sum of gauge couplings), the
identification of the third exactly marginal direction is not yet
established.

Another more recent advance in understanding these dual pairs was
the determination of the complete shortened spectrum of IIB
supergravity on $Y^{p,q}$ in \cite{Ardehali:2013a} (building on
earlier work in
\cite{Berenstein:2005xa,Kihara:2005nt,Eager:2012hx}). The shortened
supergravity multiplets are dual to protected superfields that are
all identified in the literature (see for instance
\cite{Eager:2012hx,Benvenuti:2005,Martelli:2008a,Ardehali:XX}),
\emph{with one exception: the Betti hyper multiplet} in the
supergravity spectrum.

The Betti hyper multiplet shows up in the KK spectrum because of the
non-trivial second cohomology group of $Y^{p,q}$
\cite{Eager:2012hx}. The multiplet contains a massless scalar that
is the linearized version of the gravity modulus
$\int_{\Sigma_2}B_2$ (by $\Sigma_2$ we mean the non-trivial
two-cycle in $Y^{p,q}$). This gravity modulus is dual to the third
exactly marginal deformation referred to earlier. Therefore
identifying the operator dual to the Betti hyper multiplet requires
finding the exactly marginal direction mentioned above.

In this paper we obtain a closed formula for the elusive exactly
marginal operator of $Y^{p,q}$ quivers by analyzing the NSVZ
equations. We adopt, following \cite{Imamura:2007}, the viewpoint of
brane tiling (see \cite{Franco:2006bd,Kennaway:2007} for nice
reviews of brane tilings). The authors of \cite{Imamura:2007}
realized that finding an exactly marginal deformation of field
theories described by brane tiling is equivalent to solving a linear
system of difference equations. As we show in section
\ref{sec:NSVZ}, one can solve this system of equations for $Y^{p,q}$
quivers to obtain the desired exactly marginal operator.

We will further point out that for general quivers a similar NSVZ
analysis can serve as a starting point for a systematic derivation
of protected superfields dual to supergravity Betti hyper
multiplets. In particular, for gauge theories described by brane
tiling the relevant data can be neatly encoded in a matrix (which,
for its close connection to the Konishi anomaly equation, we will
call \emph{the Konishi matrix of the quiver}) with the following two
nice properties: 1) left null vectors of the matrix yield exactly
marginal directions in the space of field theory couplings; 2) any
consistent set of baryonic charge assignments to the matter fields
gives a right null vector of the matrix.

The identification of the exactly marginal operator dual to the
$B$-field modulus of $Y^{p,q}$ theories (given in
(\ref{eq:BettiDual})) leads to a proposal for a relation between the
field theory gauge couplings and vev of the gravity modulus (given
in (\ref{eq:marginalCombination})). To check this proposal we turn
to the cascading theories obtained by adding fractional branes to
$Y^{p,q}$ geometries, and compare the change that the Reeb vector
generates in the gravity modulus, with the change in gauge theory
couplings generated by the (anomalous) U(1)$_R$; this is quite
similar to an analysis of the chiral anomaly done in
\cite{Klebanov:2002} for $T^{1,1}$.

The organization of our paper is as follows. In the next section we
review the field theoretical approach for finding exactly marginal
operators of a gauge theory described by brane tiling. In
particular, we introduce the Konishi matrix of the quivers and show
that it efficiently encodes all the data relevant for finding
exactly marginal deformations dual to $B$-field moduli. In section
\ref{sec:Ypq} we focus on $Y^{p,q}$ quivers and present a general
expression for their elusive exactly marginal operator. This leads
to a relation between the gauge theory couplings and the gravity
modulus that we further support by analyzing cascading $Y^{p,q}$
quivers. In section \ref{sec:algEx} more quiver theories are studied
using the Konishi matrix introduced in section \ref{sec:NSVZ}, and
the subtleties that may arise in cases which are more general than
$Y^{p,q}$ are discussed. Section \ref{sec:disc} contains a summary
of our results along with a few closing remarks. In appendix
\ref{app:Konishiproofs} we prove the properties of the Konishi
matrix stated in section \ref{sec:NSVZ}, and also point out how the
matrix can be thought to arise from Konishi anomaly equations
without any reference to NSVZ equations. Appendix \ref{app:Theta}
contains the field theoretical computation of the chiral anomaly of
cascading $Y^{p,q}$ quivers, and the exactly marginal directions of
the Suspended Pinch Point quiver are made explicit in appendix
\ref{app:SPP}.

\section{Exactly marginal directions in brane tilings}
\label{sec:NSVZ}

\noindent In this section we review the field theoretical approach
for finding exactly marginal operators in gauge theories described
by brane tiling. Most of the content of this chapter is already
well-known to the experts. Our only novel contribution is to
introduce the Konishi matrix in subsection
\ref{subsec:Konishimatrix}, and demonstrate its efficiency in
finding exactly marginal combinations of quiver couplings.

Brane tilings provide efficient descriptions of gauge theories on
D3-branes transverse to toric singularities
\cite{Franco:2006bd,Franco:2006}. These are quiver theories with
$N_g$ SU(N) gauge factors%
\footnote{In this paper we focus on the toric phases of the quivers;
other phases are Seiberg dual to these. See
\cite{Franco:2006bd,Franco:2006} for more details.}
with couplings $g_j$ and field strength superfields $W_{\alpha j}$,
$N_f$ chiral matter fields $\Phi_I$ in bifundamentals or adjoints of
the gauge groups, and $N_W$ superpotential terms of the form
\begin{equation*}
W_m=\pm h_m \mathrm{Tr}\prod_{I\in m}\Phi_I,
\end{equation*}
where the $\pm$ signature reflects the dimer structure of the tiling
\cite{Franco:2006bd}. Hence, $j=1,...,N_g$, $I=1,...,N_f$, and
$m=1,...,N_W$.

For all such theories \cite{Franco:2006bd}
\begin{equation}
N_g + N_W = N_f.\label{eq:Kol}
\end{equation}

In \cite{Imamura:2007}, exactly marginal combinations of gauge and
superpotential couplings were investigated from the viewpoint of
brane tiling. The authors of \cite{Imamura:2007} started with NSVZ
relations for the running of canonical gauge and superpotential
couplings
\begin{equation*}
\begin{split}
&\beta_j\equiv
\mu\frac{\mathrm{d}}{\mathrm{d}\mu}\frac{1}{g_j^2}=\frac{N/8\pi^2}{1-g_j^2
N/8\pi^2}\left[3-\frac{1}{2}\sum_{I\in j}(1-\gamma_I)\right],\\
&\beta_m\equiv\mu\frac{\mathrm{d}}{\mathrm{d}\mu}h_m=-h_m\left[3-\sum_{I\in
m}(1+\frac{\gamma_I}{2})\right].
\end{split}
\end{equation*}
In the above relations, $1+\frac{\gamma_I}{2}$ is the conformal
dimension of the chiral field $\Phi_I$, and $I\in j$ (respectively
$I\in m$) means that the chiral field $\Phi_I$ is charged under the
gauge group with coupling $g_j$ (respectively, participates in the
superpotential term with coupling $h_m$).

Next, by introducing a set of coefficients $S^a_j$ (which turn out
to play an important role in our ensuing discussions) for the gauge
couplings, and $S^k _m$ for the superpotential couplings, linear
combinations of the form
\begin{equation}
\sum_j \frac{8\pi^2 S^a _j}{N g_j '^2}-\sum_m S^k _m \log h_m,
\label{eq:ImamuraConvention}
\end{equation}
with vanishing beta functions were searched for. \emph{Any such
linear combination would correspond to an exactly marginal direction
in the space of couplings} \cite{Imamura:2007}; hence from now on we
will occasionally refer to such a set of coefficients $S^a_j,S^k_m$
as an exactly marginal direction of the field theory. The redefined
couplings $\frac{1}{g_j '^2}=\frac{1}{g _j
^2}-\frac{N}{8\pi^2}\log{\frac{1}{g _j ^2}}$, and $\log h_m$
(instead of just $h_m$) were considered because of their simpler
beta functions. Note that $1/{g'}^2$ differs from the real part of
the holomorphic coupling $1/g_{h}^2$ by $-\sum_{I\in
j}\frac{T(r_I)}{8\pi^2}\log Z_I$ \cite{Terning:2006}, where $Z_I$
denotes the wave-function renormalization of $\Phi_I$.

The authors of \cite{Imamura:2007} realized that the vanishing of
the beta function for such combinations amounts to the following
relations between the unknown coefficients $S^a_j$ and $S^k _m$
\begin{equation}
\sum_{j\in I}S^a_j=\sum_{m\in I} S^k _m\quad\text{for every }I.
\label{eq:ImamuraRelation}
\end{equation}
The left sum is over the two gauge factors under which $I$ is
charged, and the one on the right is over the two superpotential
terms in which $I$ participates. It is worth remarking that,
although it is not emphasized in \cite{Imamura:2007}, the above
relation remains true in the presence of adjoints: one only has to
count the node with adjoint twice on the LHS of
(\ref{eq:ImamuraRelation}). This is essentially because the Dynkin
index (which appears in the NSVZ beta function) for the adjoint is
twice that of the bifundamental.

To summarize this section so far, the problem of finding exactly
marginal deformations of the gauge and superpotential couplings of a
CFT described by brane tiling is reduced to finding a set of
coefficients ${S^a_j,S^k_m}$ that solve (\ref{eq:ImamuraRelation});
these yield RG invariant combinations of the form
(\ref{eq:ImamuraConvention}). We conjecture that the corresponding
exactly marginal operators are
\begin{equation}
\sum_j S^a_j (\mathrm{Tr} W_j^2)-\sum_m S^k_m
(\frac{32\pi^2}{N}W_m). \label{eq:primaryMarginal1}
\end{equation}
Evidence for this conjecture will be presented after equation
(\ref{eq:Y84BettiDual}), and also in appendix
\ref{app:Konishiproofs}.

In \cite{Imamura:2007}, two sets of solutions to
(\ref{eq:ImamuraRelation}) were found for an arbitrary gauge theory
described by brane tiling.
\begin{itemize}
\item The first solution was $S^a_j =S^k _m =1$ for all $j$ and $m$;
this direction in the space of couplings corresponds to \textbf{the
sum of the (inverse squared of) gauge couplings}, and is dual to the
supergravity \emph{axion-dilaton}.
\item The second solution was $S^a _j =0$ for all $j$, and $S^k _m
=\pm1$, with the signature depending on the sign of the
superpotential term $m$; this is \textbf{the $\beta$-deformation} of
the field theory, with \emph{Lunin-Maldacena} gravity dual \cite{Lunin:2005}.\\
\end{itemize}

In the present paper we are interested in going beyond the two sets
of general solutions mentioned above. It is well-known (and we will
review shortly) that if the quiver CFTs under study are dual to
SE$_5$ geometries with non-trivial two-cycles,
\begin{itemize}
\item there are additional solutions to (\ref{eq:ImamuraRelation})
which describe exactly marginal directions arising from baryonic
symmetries in field theory. We will refer to these as
\textbf{B-deformations}. These are dual to \emph{the B-field moduli
on the non-trivial two-cycles}.
\end{itemize}

In particular, $Y^{p,q}$ quivers have one such exactly marginal
direction, as we will describe in the next section.

We now explain how such additional exactly marginal directions
originate from baryonic symmetries of the field theory.

\subsection{Exactly marginal directions and baryonic charge assignments}
\label{subsec:baryons}

\noindent In this subsection we review the argument of
\cite{Martelli:2008a} demonstrating the appearance of additional
exactly marginal directions when global baryonic U(1) symmetries are
present.

Recall that one can imagine that the $N_g$ SU(N) gauge groups in the
IR, started out as U(N) gauge groups in the UV, for D3-branes
probing a cone over SE$_5$. However, all the U(1)'s decouple in the
IR: the `center of mass' U(1) decouples as nothing is charged under
it; $b_3 ($SE$_5)$ (defined as the rank of the third homology group
of SE$_5$) of the massless U(1)'s decouple because their gauge
coupling goes to zero in the infrared, but they nevertheless yield
non-anomalous global baryonic symmetries in the IR theory; the
remaining $N_g - b_3 (\text{SE}_5) -1$ U(1)'s become massive and
yield anomalous baryonic currents in the IR.

We are interested in the $b_3 ($SE$_5)$ non-anomalous baryonic
symmetries. For each of these, we have a baryonic charge assignment
for the chiral fields. The baryonic U(1) charges $Q_J$ (with
$J=1,...,b_3 ($SE$_5)$) must satisfy (see for example
\cite{Herzog:2003})
\begin{equation}
\begin{split}
\sum_{I\in j} Q_J (\Phi_I)=0 \quad &\text{for every node }j,\\
Q_J(\Phi_{j_1\ j_2})+Q_J(\Phi_{j_2\ j_3})+...+Q_J(\Phi_{j_r\ j_1})=0
\quad &\text{for every loop}.\label{eq:bChargeCond}
\end{split}
\end{equation}
Note that the nodes and loops referred to in the above relations are
the ones in the quiver diagram picture, not in the brane tiling.

Now, as argued in \cite{Martelli:2008a}, for any $R$-charge
assignment $\{R^{\ast}_I\}$ to the chiral fields yielding zero beta
functions for the couplings, the relations in (\ref{eq:bChargeCond})
guarantee that $\{R^{\ast}_I+\mu Q_J (\Phi_I)\}$, for any
$\mu\in\mathbb{R}$, is another zero of the beta functions. Thus, for
each one of the $b_3 ($SE$_5)$ baryonic charge assignments to the
chiral fields there exists an exactly marginal deformation of the
theory. We referred to these as B-deformations. The interested
reader can find another field theoretical explanation for the origin
of B-deformations in baryonic symmetries from the viewpoint of the
Konishi anomaly equation in appendix \ref{app:Konishiproofs}.

The gravity analog of the above field theoretical analysis is as
follows. Each non-trivial two-cycle in SE$_5$ yields a massless
Betti vector in the supergravity KK spectrum. This massless vector
is dual to a conserved baryonic current. On the other hand, the
non-trivial two-cycle yields a gravity modulus of the
form\footnote{At the linearized level the gravity modulus is a
component of a Betti hyper multiplet in the bulk KK spectrum, and it
is not difficult to see why it remains massless at nonlinear level
too: since Betti multiplets are singlet under the isometry group,
turning them on does not Higgs any gauge symmetries in the bulk, and
hence they are exactly marginal. See footnote 4 for a similar
argument in more detail.}
$\int_{\mbox{\footnotesize{2-cycle}}}B_{2}$. This gravity modulus is
dual to an exactly marginal operator. Hence, the one-to-one
correspondence between baryonic symmetries and their related exactly
marginal deformations.

To summarize, there are $b_3($SE$_5)+2$ solutions to
(\ref{eq:ImamuraRelation}). $b_3($SE$_5)$ of them correspond to
B-deformations, and arise from baryonic symmetries. The remaining
two correspond to the $\beta$-deformation and the axion-dilaton;
these can be thought to arise from the global U(1)$\times$U(1)
flavor symmetry (see the Konishi anomaly discussion in appendix
\ref{app:Konishiproofs}).

It seems like in general a useful rule of thumb is that additional
global symmetries in field theory are responsible for a larger
conformal manifold. In such cases as $Y^{p,q}$ quivers, we saw that
the global baryonic symmetry dual to the non-trivial two-cycle
explains the additional exactly marginal direction. We will come
back to this rule of thumb a few times.

\subsection{A simple algorithm for toric quivers}
\label{subsec:Konishimatrix}

\noindent In this section, we provide a simple algorithm for finding
exactly marginal combinations of the form
\begin{equation}
\sum_j \frac{8\pi^2 S^a _j}{N g_j '^2}-\sum_m S^k _m \log h_m,
\label{eq:ImamuraConventionRep}
\end{equation}
for quiver theories described by brane tiling. This algorithm is an
immediate consequence of the work in \cite{Imamura:2007}.

The central ingredient of the algorithm is a neatly derivable
matrix, $B_{K}$, that we will refer to as the Konishi matrix of the
quiver. The columns of $B_{K}$ are labeled by the chiral fields
$I=1,...,N_f$. The first $N_g$ rows of the matrix are labeled by the
gauge groups $j=1,...,N_g$. The rest of the rows are labeled by the
superpotential terms $m=1,...,N_W$. Equation (\ref{eq:Kol}) implies
that $B_{K}$ is an $N_f \times N_f$ square matrix for quivers
described by brane tiling.

The entries of the Konishi matrix are filled, in a rather natural
way, as follows. For a row labeled by a gauge group $j$ insert one
in column $I$ if $I\in j$ (insert two if $I$ is in the adjoint of
the $j$th gauge group), and insert zero otherwise. For a row labeled
by a superpotential term $m$ insert $-1$ in column $I$ if $I\in m$,
and zero otherwise.

Note that the direction of the bifundamental chiral field does not
matter in constructing the Konishi matrix. This is because the
Dynkin index appearing in the NSVZ beta function is quadratic in the
generators.

Now that we described how to form $B_{K}$, we explain two of its
main properties. The proofs can be found in appendix
\ref{app:Konishiproofs}.
\begin{itemize}
\item Every baryonic charge assignment satisfying (\ref{eq:bChargeCond})
gives an $N_f$ tuple $(Q_J(X_1),...,Q_J(X_{N_f}))^T$ that is a null
vector of $B_{K}$.
\item Every marginal direction of the form (\ref{eq:ImamuraConventionRep}) in the space of couplings gives
an $N_f$ tuple $(S^a_1 ,...,S^a_{N_g},S^k_1,...,S^k_{N_W})$ that is
a left null vector of $B_{K}$.
\end{itemize}

In other words, baryonic charges form null vectors of $B_{K}$, while
exactly marginal combinations give null vectors of $B^T_{K}$.

For the exactly marginal directions, the statement goes the other
way as well: every null vector of $B^T_{K}$ yields an exactly
marginal combination of the form (\ref{eq:ImamuraConventionRep}).
However, not every null vector of $B_{K}$ gives a consistent
baryonic charge assignment. The reason is that the second condition
in (\ref{eq:bChargeCond}) is not fully ensured for the null vectors
of $B_{K}$; only the loops that appear in the superpotential are
taken into account by $B_{K}$. As we explain in appendix
\ref{app:Konishiproofs}, the remaining relevant data for
constraining baryonic charge assignments is contained in the two
non-trivial cycles of the torus of brane tiling. In fact, since the
$\beta$-deformation and the sum of gauge couplings are two exactly
marginal directions that do not correspond to any baryonic
symmetries, we did expect that the non-trivial baryonic charge
assignments be two fewer than the exactly marginal deformations. The
previous statements are made more precise in appendix
\ref{app:Konishiproofs}.

The question arises: how can we then determine the codimension two
subspace of the null space of $B_{K}$ that corresponds to the
non-trivial baryonic charge assignments?

The answer follows from the statement \cite{Imamura:2007} that
$S^a_{t(I)}-S^a_{h(I)}$ (with a hopefully obvious notation for
`head' and `tail' of a chiral field) is a consistent assignment of
baryonic charge to $\Phi_I$. This charge would be zero for all
$\Phi_I$ if we take the $S^a$ coefficients of the two general
solutions corresponding to the $\beta$-deformation (with all $S^a$
equal to zero) and the axion-dilaton (with all $S^a$ equal); the
remaining $b_3 ($SE$_5)$ exactly marginal directions are the ones
that correspond to non-trivial baryonic charge assignments.

Now, when $b_3 ($SE$_5) >1$, there are more than one B-deformations
in the field theory, and one would like to be able to put these in a
one-to-one correspondence with the non-trivial two-cycles in the
dual geometry. This is where the knowledge of baryonic charge
assignments corresponding to each two-cycle in the dual geometry
becomes necessary. Such charge assignments can be algorithmically
derived for toric quivers as explained in \cite{Franco:2006}. Then
comparing $S^a_{t(I)}-S^a_{h(I)}$ of the exactly marginal directions
with the baryonic charge assignments $Q_J(X_I)$ of the two-cycles
yields the correspondence between the exactly marginal deformations
and their related non-trivial two-cycles in the dual geometry. An
example of this kind will be considered in section \ref{sec:algEx}.

To illustrate the above algorithm we now study exactly marginal
directions of the Klebanov-Witten CFT dual to IIB theory on
AdS$_5\times T^{1,1}$ \cite{Klebanov:1998}. The Konishi matrix for
this theory is
\begin{equation*}
B_{K}=
\begin{pmatrix}
  1 & 1 & 1 & 1 \\
  1 & 1 & 1 & 1 \\
  -1 & -1 & -1 & -1\\
  -1 & -1 & -1 & -1
 \end{pmatrix}.
\end{equation*}
Recall that the first two rows correspond to the two gauge factors,
the last two rows to the superpotential terms, and the columns to
the four bifundamental chiral fields.

The left null vectors of the above matrix are
$(S^a_1,S^a_2,S^k_1,S^k_2)=(1,1,1,1)$, $(0,0,1,-1)$, and
$(1,-1,0,0)$. The first one corresponds to the sum of the gauge
couplings, the second one to the $\beta$-deformation, and the third
one to the B-deformation (the difference of the gauge couplings).
These exactly marginal directions are all well known
\cite{Klebanov:1998,Benvenuti:2005a}. In fact, more exactly marginal
directions are known for this theory \cite{Benvenuti:2005a} than the
three we mentioned. There exist two extra exactly marginal
deformations of the theory that arise from adding mesonic exactly
marginal operators to the superpotential that were not present in
the superpotential of the original theory. These were called
``accidentally marginal'' operators in \cite{Imamura:2007}. Such
accidentally marginal deformations are (except for a few remarks)
completely ignored in our paper; we only consider deformations by
changing the couplings already present in the original theory.
Before moving on, however, we would like to point out that the
existence of these extra exactly marginal deformations for the
conifold theory is consistent with the rule of thumb we mentioned
earlier: a bigger global symmetry group yields a larger conformal
manifold. In this case, the bigger symmetry group of the conifold
theory (compared to $Y^{p,q}$ quivers) forbids the presence of all
the exactly marginal operators in the superpotential, but these
extra operators could serve to deform the theory later.

To summarize, we presented an algorithm for finding the coefficients
$S^a$ and $S^k$ related to the marginal directions from the
$N_f\times N_f$ Konishi matrix of the quiver; from knowledge of
$S^a$ coefficients one can then recover the baryonic charge
assignments via $S^a_{t(I)}-S^a_{h(I)}=Q_J(X_I)$. We should point
out that there already exists a simpler algorithm in the literature
for finding only the $S^a$ coefficients (and thereby, the baryonic
charge assignments) from the $N_g\times N_g$ `incidence matrix' of
the quiver (see for example \cite{Kennaway:2007}). The incidence
matrix is smaller than the Konishi matrix, and therefore easier to
compute with, but it can not yield the exactly marginal directions.
The extra information that the Konishi matrix is capable to give us
is the coefficients $S^k$, which along with $S^a$ serve to fully
specify the exactly marginal directions. Finally, the reader should
note that it is not possible to obtain the ``accidentally marginal''
directions (which arise from mesonic $R$-charge 2 chiral primaries
absent in the superpotential) from the Konishi matrix; finding such
deformations would require a separate analysis which is not covered
at all in the present paper.

We will illustrate the above algorithm with more explicit examples
in section \ref{sec:algEx}.

\section{The B-deformation of $Y^{p,q}$ quivers}
\label{sec:Ypq}

\noindent In this section we find the exactly marginal deformation
of $Y^{p,q}$ quivers which is dual to the $B$-field modulus on the
gravity side. This is done by directly solving equations
(\ref{eq:ImamuraRelation}) for $Y^{p,q}$ quivers; no use of the
Konishi matrix is made in this section. Based on our result we then
propose a relation between the gauge theory couplings and the vev of
the complex $B$ field on the non-trivial two-cycle of $Y^{p,q}$. The
proposal is further supported by considering the effect of adding
fractional branes to $Y^{p,q}$ geometries.

The superfield version of the exactly marginal operator that we find
is dual to the Betti hyper multiplet in the supergravity KK
spectrum. This result incidentally completes the identification of
the protected operators dual to shortened supergravity KK multiplets
on generic $Y^{p,q}$. Therefore, we see it appropriate to start by
reviewing the light multiplets of supergravity and their
low-dimension dual operators.

\subsection{AdS/CFT state-operator correspondence for $Y^{p,q}$}
\label{subsec:stateop}

\noindent The shortened (protected) spectrum of IIB supergravity on
AdS$_5\times Y^{p,q}$ is detailed in \cite{Ardehali:2013a}. There
are nine towers of supermultiplets called Graviton, Gravitino 1--4
and Vector 1--4, each filled with representations of SU(2,2$|$1),
that we denote by $\mathcal{D}(E_0,s_1,s_2;r)$. Each bulk multiplet
also transforms under a specific representation of the isometry
group of $Y^{p,q}$, which is SU(2)$_j \times$U(1)$_{\alpha}
\times$U(1)$_R$; we label representations of the SU(2) by $j$ and
$-j\le N_{\phi}\le j$, representations of the U(1)$_{\alpha}$ by
$N_{\alpha}$, and representations of the U(1)$_R$ by $N_{\psi}$. See
\cite{Ardehali:2013a} for the detailed quantization conditions.

\subsubsection*{Conserved multiplets}

We begin the discussion with conserved multiplets in the spectrum.
There is one conserved multiplet in the Graviton tower. It contains
the massless graviton dual to the boundary stress tensor, and a
massless vector dual to the boundary $R$-current.

The Vector 1 tower generically contains four conserved vectors, a
$j=1$ triplet with $N_{\psi}=N_{\alpha}=0$ and a $j=0$ singlet with
$N_{\psi}=N_{\alpha}=0$; these are dual to the boundary flavor
currents, in the triplet of the SU(2) and the singlet of the U(1) of
the isometry group of $Y^{p,q}$
\begin{equation*}
J_{SU(2)_j}^{k}=\sum_{i}\mbox{Tr }U_i
e^{V_{h(i)}}\sigma^k\bar{U_i}e^{-V_{t(i)}}+\sum_{i}\mbox{Tr }V_i
e^{V_{h(i)}}\sigma^k\bar{V_i}e^{-V_{t(i)}},
\end{equation*}
\begin{equation}
J_{U(1)_{\alpha}}=\sum_{i}\mbox{Tr }V_i
e^{V_{h(i)}}\bar{V_i}e^{-V_{t(i)}}-\sum_{i}\mbox{Tr }Y_i
e^{V_{h(i)}}\bar{Y_i}e^{-V_{t(i)}}+\sum_{i}\mbox{Tr }Z_i
e^{V_{h(i)}}\bar{Z_i}e^{-V_{t(i)}},\label{eq:U1current}
\end{equation}
where the sums are over the bifundamental chiral superfields ($U_i ,
V_i , Y_i , Z_i$) in the quiver, $\sigma^k$ are the Pauli matrices,
and $V_{h(i)}$ (or $V_{t(i)}$) in each term is the Lie algebra
valued superfield of the gauge factor at the head (or tail) of the
corresponding chiral bifundamental. We hasten to remind the reader
that $Y^{p,q}$ quivers consist of $2p$ nodes and a number of chiral
bifundamentals of doublet type ($U,V$) and singlet type ($Y,Z$); see
Figure \ref{fig:Y84} for an example, or \cite{Benvenuti:2004} for a
review. The coefficients of the terms on the RHS of
(\ref{eq:U1current}) reflect the U(1)$_{\alpha}$ charges of the
chiral fields.

In the previous paragraph we emphasized generically, because
$Y^{1,0}$ and $Y^{2,0}$ (also known as $T^{1,1}$ and
$T^{1,1}/\mathbb{Z}_2$ respectively) are exceptional: in these
geometries there are six such conserved vector multiplets, half of
them in the triplet of one SU(2) and the other half in the triplet
of the other SU(2) of their isometry group SU(2)$_j \times$SU(2)$_l
\times$U(1)$_R$.

The other conserved multiplet in the bulk is the Betti-vector
multiplet which has zero $R$-charge and is singlet under the
isometry group of $Y^{p,q}$. The dual boundary operator is a
baryonic superfield of the schematic form \cite{Martelli:2008a}
\begin{equation*}
\begin{split}
\mathcal{U^I}=&-p\sum_{i}\mbox{Tr }U_i
e^{V_{h(i)}}\bar{U_i}e^{-V_{t(i)}}+q\sum_{i}\mbox{Tr }V_i
e^{V_{h(i)}}\bar{V_i}e^{-V_{t(i)}}\\
&+(p-q)\sum_{i}\mbox{Tr }Y_i
e^{V_{h(i)}}\bar{Y_i}e^{-V_{t(i)}}+(p+q)\sum_{i}\mbox{Tr }Z_i
e^{V_{h(i)}}\bar{Z_i}e^{-V_{t(i)}}.\label{eq:baryonCurrent}
\end{split}
\end{equation*}

There are no more conserved multiplets in the spectrum, unless
$p=q=1$, where one has $Y^{1,1}=S^5/\mathbb{Z}_2$ with enhanced
isometry and supersymmetry, hence additional conserved gravitino
multiplets.

\subsubsection*{Chiral multiplets with massless scalar
components}

We now discuss the bulk chiral multiplets with massless scalar
components. These scalars are the duals of marginal single-trace
deformations at the linearized level.

First, there is the universal hyper multiplet in the Vector 4 tower
(and its CP conjugate in Vector 3), transforming in
$\mathcal{D}(3,0,0;2)$ (and $\mathcal{D}(3,0,0;-2)$) of SU(2,2$|$1).
This multiplet is a singlet under the internal isometry (it has
$j=N_{\psi}=N_{\alpha}=0$) and is dual to\footnote{As in
(\ref{eq:primaryMarginal1}), there are correction terms proportional
to $S^k_m W_m$ that we are dropping for convenience.}
$\mbox{Tr}W^{\alpha}W_{\alpha}$. The massless scalar inside this
multiplet comes from the ten dimensional axion-dilaton and can be
identified with the sum of the holomorphic gauge couplings in the
dual quiver $\sum_i \tau_i$; this is the first modulus.

Next, there are generically three chiral multiplets (and their CP
conjugates) in Vector 1 transforming in $\mathcal{D}(3,0,0;2)$ (and
$\mathcal{D}(3,0,0;-2)$) of SU(2,2$|$1), that transform as a triplet
of the SU(2) of the isometry group of $Y^{p,q}$ with $j=N_{\psi}=1$
and $N_{\alpha}=0$; this triplet is dual to the $\mathcal{S}$ meson
of the field theory \cite{Benvenuti:2005}. Again we emphasized
generically, because for $Y^{1,0}$ and $Y^{2,0}$ there are nine such
multiplets, transforming in the triplet of both SU(2) factors in
their isometry group. These multiplets contain massless scalars
dual, at the linearized level, to the $\beta$-deformation (or in the
cases of $Y^{1,0}$ and $Y^{2,0}$, also to the PW and $\lambda_2$
``accidentally marginal'' deformations \cite{Benvenuti:2005a}) of
the field theory. Thus, for $Y^{1,0}$ and $Y^{2,0}$ three out of the
nine massless scalars are actual moduli, and for other $Y^{p,q}$ one
out of these three massless scalars is\footnote{This can be easily
seen from an argument following \cite{Green:2010}: upon turning on
the triplet of scalars, the SU(2) of the internal isometry breaks
down to U(1), thus the two massless vectors in the bulk that used to
gauge the broken SU(2)/U(1) need to become massive by eating two of
the formerly massless scalars and making them massive too; only one
remains massless at nonlinear level.}; this is the second modulus.

Finally, and of most interest to us, there is the singlet
Betti-hyper multiplet, and its CP conjugate, transforming in
$\mathcal{D}(3,0,0;2)$, and $\mathcal{D}(3,0,0;-2)$, of SU(2,2$|$1).
These have as their scalar component the vev of the complex $B$
field on the two-cycle of $Y^{p,q}$; this is the third and last
modulus\footnote{Exceptional cases $Y^{1,1}$, $Y^{2,2}$, and
$Y^{3,3}$ have extra massless scalars in their shortened
supergravity spectrum that we do not consider in the current work.}.

Our main proposal in this section is that the Betti hyper multiplet
is dual to the following operator on the field theory side
\begin{eqnarray}
\mathcal{B^I}=&&\sum_{j=1}^{p-q} [(-1)^{p-q+j}(p+\frac{q}{2})+q j-\frac{q}{2}](\mbox{Tr }W_{j}^2-\mbox{Tr }W_{j'}^2)\nn\\
&&+\sum_{j=p-q+1}^{p}
[(-1)^{p-q+j}\frac{p+q}{2}-j(p-q)+(p+\frac{1}{2})(p-q)](\mbox{Tr
}W_{j}^2-\mbox{Tr }W_{j'}^2).\label{eq:BettiDual}
\end{eqnarray}
The numbering of the nodes is explained in the next subsection. The
nodes $1,1',\ldots,p-q,(p-q)'$ have $U,Y,Z$ chiral bifundamentals
attached to them and will be referred to as `impurity' nodes, while
the rest have $U,V,Y,Y$ attached to them and will be called `clean'
nodes; $Y^{p,p}$ quivers are completely clean. Similar to
(\ref{eq:primaryMarginal1}), one should add correction terms
proportional to $S^k_m W_m$ on the RHS of (\ref{eq:BettiDual}) that
we have suppressed for convenience.

Let us look at a couple of examples. For $Y^{1,0}=T^{1,1}$ this
operator takes the expected form
\begin{equation}
\mathcal{B^I}(T^{1,1})=\mbox{Tr }[W_1 ^2 -
W_{1'}^2].\label{eq:BettiDualT11}
\end{equation}
This was called the `exceptional chiral operator' in
\cite{Baumann:2010}, as it does not belong to any tower of protected
single-trace operators (similarly on the gravity side the Betti
hyper multiplet does not belong to any KK tower). One can see from
the above expression that the difference of (inverse squared) gauge
couplings is the field theory dual of the vev of the gravity modulus
inside the Betti-hyper on $T^{1,1}$. We explained, at the end of the
previous section, how an NSVZ analysis leads to the exact
marginality of this combination. Note that since in this case $S^k$
coefficients are zero, no correction terms should be added to the
RHS of (\ref{eq:BettiDualT11}), and hence no tuning of
superpotential couplings is required for this B-deformation.

As another example, for $Y^{1,1}=S^5/\mathbb{Z}_2$ the corresponding
operator takes the form
\begin{equation}
\mathcal{B^I}(S^5/\mathbb{Z}_2)=-\mbox{Tr }[W_1 ^2 -
W_{1'}^2].\label{eq:BettiDualZ2orb}
\end{equation}
(The different sign for the first two terms, as compared to
(\ref{eq:BettiDualT11}), is only due to our convention in numbering
the nodes; see Figure \ref{fig:Y84} for an explanation.) This
operator is in the twisted sector of the field theory. The dual
Betti-hyper multiplet is identified in the twisted sector of IIB
theory compactified on $S^{5}/\mathbb{Z}_2$ \cite{Ardehali:2013}. In
this case $S^k$ coefficients turn out to be nonzero, and correction
terms proportional to $S^k_m W_m$ should be added to the operator.

From (\ref{eq:BettiDual}) we claim that the vev of the complex $B_2$
field on the two-cycle of $Y^{p,q}$ is related to the gauge
couplings of the dual quiver in the following way\footnote{See our
comments after (\ref{eq:Y84BettiDual}) for a partial reasoning
behind our proposals in (\ref{eq:BettiDual}) and
(\ref{eq:marginalCombination}). Also, as explained in footnote 8,
the following equation neglects the non-zero value of the $B$ field
at the point where all the gauge theory couplings are equal. Other
than that, our conventions are the same as in \cite{Herzog:2001r}.}
\begin{eqnarray}
\frac{1}{2\pi^2\alpha'}\int_{\Sigma_2} (C_2-i\frac{B_2}{g_s}) =&&
\sum_{j=1}^{p-q} [(-1)^{p-q+j}(p+\frac{q}{2})+q j-\frac{q}{2}]
(\tau_{j}-\tau_{j'})\nn\\
&&+\sum_{j=p-q+1}^{p}
[(-1)^{p-q+j}\frac{p+q}{2}-j(p-q)+(p+\frac{1}{2})(p-q)](\tau_{j}-\tau_{j'}),\nn\\
\label{eq:marginalCombination}
\end{eqnarray}
where $\tau_{j}=\frac{\Theta_j}{2\pi}+\frac{4\pi i}{g_j ^2}$ are the
holomorphic gauge couplings.

In the rest of this section we are going to support the above
proposal, first by outlining how the appearing coefficients solve
the appropriate NSVZ relations, and then by showing that the
proposal is correct in a background with added fractional branes.

\subsection{The marginal direction from NSVZ}

\begin{figure}[t]
\centering
    \includegraphics[scale=.9]{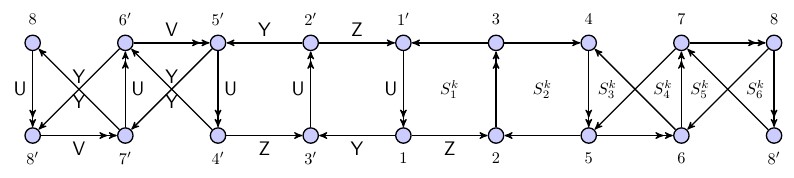}
\caption{The quiver for $Y^{8,4}$ is shown to demonstrate the way we
have numbered the nodes and faces of $Y^{p,q}$ quivers in general.
The quivers are formed from impurity blocks with $Z$ or clean blocks
with $V$ in them (see \cite{Herzog:2005}). We draw the quiver in the
most parity symmetric way with the impurity blocks in the middle.
For even $p-q$, assign number 1 to the node in the middle with $Z$
leaving it, and the numbers increase along $Z$ and $U$
bifundamentals until node $p$ is arrived at. Note that for every
numbered node in the quiver there is a mirror node that we denote
with a prime. $S^k_j$ denote the coefficients of the superpotential
couplings in (\ref{eq:ImamuraConvention}); for every $S^k$ there is
a mirror coefficient $S'^k=-S^k$ on the left that we have not shown
in the figure. For odd $p-q$ there will be a single impurity block
in the middle with $S^k_0=0$. Then number 1 is assigned to the node
in the middle with $Z$ entering it. \label{fig:Y84}}
\end{figure}

\noindent Let us start by listing the couplings of $Y^{p,q}$ quiver
theories; see \cite{Benvenuti:2004} for a detailed review. First,
there are $2p$ gauge couplings $g_i$, one for each node. Next, there
are $2p-2q$ superpotential couplings $h_m$, two for each square face
(see Figure \ref{fig:Y84}), that multiply quartic terms of the form
\begin{equation*}
Z_{j+1\ j+2}U^{1}_{j+2\ j+3} Y_{j+3\ j} U^{2}_{j\
j+1},\quad\mbox{or}\quad Z_{j+1\ j+2}U^{2}_{j+2\ j+3} Y_{j+3\ j}
U^{1}_{j\ j+1}.
\end{equation*}
Finally there are $4q$ superpotential couplings $h_{m}$ (with $m$
different from those of quartic couplings), two for each triangular
face, that multiply cubic terms of the form
\begin{equation*}
\begin{split}
&U^{1}_{j\ j+1} V^2_{j+1\ j+2} Y_{j+2\ j},\quad\mbox{or}\quad
U^{2}_{j\ j+1} V^1_{j+1\ j+2} Y_{j+2\ j},\\
\mbox{or}&\quad V^{1}_{j\ j+1} U^2_{j+1\ j+2} Y_{j+2\
j},\quad\mbox{or}\quad V^{2}_{j\ j+1} U^1_{j+1\ j+2} Y_{j+2\ j}.
\end{split}
\end{equation*}

Now, as explained in the previous section, one can look for linear
combinations of gauge and superpotential couplings (of the form
shown in (\ref{eq:ImamuraConvention})) that have vanishing beta
functions. Such combinations have coefficients $S^a$ and $S^k$ that
satisfy (\ref{eq:ImamuraRelation}). Since $b_3 (Y^{p,q})$ is
one\footnote{Recall that all $Y^{p,q}$ manifolds, with $p> q\ge 0$,
are smooth and are topologically $S^2\times S^3$. The special cases
of $Y^{p,p}$ (also known as $S^5 /\mathbb{Z}_{2p}$) are not smooth,
but when their fixed circle is blown up they also acquire the
topology $S^2\times S^3$ and hence a third betti number equal to
one.}, there is precisely one B-deformation in the space of
couplings of any $Y^{p,q}$ quiver, with $p\ge q\ge 0$.

The $S^a$ and $S^k$ coefficients that characterize the B-deformation
of these theories are:
\begin{equation}
\begin{split}
&S^{a}_j=\begin{cases}(-1)^{p-q+j} (p+\frac{q}{2})+ q j-\frac{q}{2}\quad &\mbox{for}\quad 1\le j\le p-q\quad\mbox{(impurity nodes)},\\
(-1)^{p-q+j} \frac{p+q}{2}-j(p-q)+(p+\frac{1}{2})(p-q)\quad
&\mbox{for}\quad p-q+1\le j\le p\quad\mbox{(clean
nodes)},\end{cases}\\
&S^{k}_{j}=\begin{cases}(2j-\frac{(-1)^{p-q}+1}{2})q \hspace{2.2cm}\mbox{for}\quad 1\le j\le \lfloor\frac{p-q}{2}\rfloor\quad\mbox{(quartic faces)},\\
-(-1)^{j-\lfloor\frac{p-q}{2}\rfloor}
\frac{p+q}{2}-(j-\lfloor\frac{p-q}{2}\rfloor)(p-q)+(p-q)(q+\frac{1}{2})\\
\hspace{5.1cm}\mbox{for}\quad \lfloor\frac{p-q}{2}\rfloor+1\le j\le
\lfloor\frac{p-q}{2}\rfloor+q\quad\mbox{(cubic faces)}.\end{cases}\\
\end{split}
\label{eq:margCos}
\end{equation}
The numbering is explained in Figure \ref{fig:Y84}. Also, only half
of the $S$ coefficients are presented in (\ref{eq:margCos}); the
other half mirror the above set, but come with the sign flipped.

Note that despite every face of the quiver yielding two
superpotential terms, we have assigned only one $S^k$ coefficient to
each face. This is because we are looking for a solution that does
not break the global SU(2) symmetry. So every face does come with
two $S^k$ coefficients, but the two are equal for our solution. This
would clearly not be true if one considered SU(2) breaking solutions
such as the one corresponding to the $\beta$-deformation.

As an example, we present the coefficients one obtains from
(\ref{eq:margCos}) for the case $p=8,q=4$ shown in Figure
\ref{fig:Y84}:
\begin{equation*}
\begin{split}
&S^a_1=-S^a_{1'}=-8,\ S^a_2=-S^a_{2'}=16,\ S^a_3=-S^a_{3'}=0,\ S^a_4=-S^a_{4'}=24,\\
&S^a_5=-S^a_{5'}=8,\ S^a_6=-S^a_{6'}=16,\ S^a_7=-S^a_{7'}=0,\ S^a_8=-S^a_{8'}=8,\\
&S^k_1=-S^k_{1'}=4,\ S^k_2=-S^k_{2'}=12,\\
&S^k_3=-S^k_{3'}=20,\ S^k_4=-S^k_{4'}=4,\ S^k_5=-S^k_{5'}=12,\
S^k_6=-S^k_{6'}=-4.
\end{split}
\end{equation*}
In particular, from the above coefficients the following expression
for $\mathcal{B^I}_{Y^{8,4}}$ is obtained
\begin{equation}
\begin{split}
\mathcal{B^I}_{Y^{8,4}}=&-8(\mbox{Tr }W_{1}^2-\mbox{Tr
}W_{1'}^2)+16(\mbox{Tr }W_{2}^2-\mbox{Tr }W_{2'}^2)+24(\mbox{Tr
}W_{4}^2-\mbox{Tr }W_{4'}^2)\\
&+8(\mbox{Tr }W_{5}^2-\mbox{Tr }W_{5'}^2)+16(\mbox{Tr
}W_{6}^2-\mbox{Tr }W_{6'}^2)+8(\mbox{Tr }W_{8}^2-\mbox{Tr
}W_{8'}^2).\label{eq:Y84BettiDual}
\end{split}
\end{equation}

Similarly, it is the expression for the $S^a$ coefficients in
(\ref{eq:margCos}) that has led us to propose (\ref{eq:BettiDual})
for general $Y^{p,q}$. Our main reason for proposing
(\ref{eq:BettiDual}) is that, as explained above, it yields the
expected forms in the special cases with $p=1$. Further partial
support comes from the analysis of the Konishi anomaly equation in
appendix \ref{app:Konishiproofs}. Also, with (\ref{eq:BettiDual}) at
hand it seems natural to expect (\ref{eq:marginalCombination}), and
the latter equation will be supported in subsection
\ref{subsec:ChiAnom}. The way one is led to
(\ref{eq:marginalCombination}) from (\ref{eq:BettiDual}) is as
follows. Take the $Y^{1,1}=S^5/\mathbb{Z}_2$ example. If we assume
that the undeformed theory has equal couplings for the two gauge
factors and its gravity dual has zero vev for the complex $B_2$
field\footnote{This assumption is in fact incorrect. The vev of the
complex $B_2$ field is non-zero when the gauge couplings are equal;
see \cite{Lawrence:1998} for the $Y^{1,1}$ case, and equation (19)
in \cite{Herzog:2001r} for $Y^{1,0}$. However, our argument goes
through, and our result is correct, up to this non-zero additive
constant.}, then turning on the vev would be dual to turning on the
deformation operator (\ref{eq:BettiDualZ2orb}). Therefore the vev in
the deformed gravity theory would be proportional to the difference
of the gauge couplings in the deformed field theory. That the
proportionality constant in (\ref{eq:marginalCombination}) is
correct will be argued in subsection \ref{subsec:ChiAnom}. Note that
the $S^k$ coefficients, if nonzero, would signal the required tuning
of the superpotential couplings in the deformation, but do not enter
(\ref{eq:marginalCombination}) themselves.

Rather than proving the relations in (\ref{eq:margCos}) we
demonstrate their validity, and in fact only partially; the
interested reader can complete the analysis along similar lines.
Consider a case where $p-q$ is even, so that the quiver looks like
Figure \ref{fig:Y84}. Take two nodes numbered $2l+1$ and $2l+2\le
p-q$ according to the procedure explained under Figure
\ref{fig:Y84}. These are connected by a $Z$ chiral bifundamental
that participates in two quartic superpotential terms that enter in
the deformation with coefficient $S^k _{l+1}$. Equation
(\ref{eq:ImamuraRelation}) applied to the bifundamental superfield
$Z$ reads
\begin{equation*}
S^a_{2l+1}+S^a_{2l+2}=2S^k _{l+1}.
\end{equation*}
It is easy to check that this equation is satisfied by the
coefficients in (\ref{eq:margCos}) since we have
$S^a_{2l+1}=-p+2lq$, $S^a_{2l+2}=p+(2l+2)q$, and $S^k
_{l+1}=(2l+1)q$. Similar computations can demonstrate the full
validity of (\ref{eq:margCos}).

We take a moment to remind the reader that if one wanted to find
only the $S^a$ coefficients, they would have an easier job since the
$S^a$ coefficients can be obtained from the knowledge of the
baryonic charges via ($S^a_{t(I)}-S^a_{h(I)}=Q_J (X_I)$). But to
find the $S^k$ coefficients as well, one should solve
(\ref{eq:ImamuraRelation}).

\subsection*{Important features of the solution}

\noindent The first important feature of the solution in
(\ref{eq:margCos}) that we would like to point out is that unless
$q=0$ (corresponding to orbifolds of $T^{1,1}$) the coefficients
$S^k$ are non-zero; this means that moving along this marginal
direction requires not only changing a linear combination of the
(inverse squared of the) gauge couplings, but also tuning the
superpotential couplings in an appropriate way. This was pointed out
in \cite{Benvenuti:2005a}.

The coefficients with which the gauge couplings appear in this
marginal combination are of most importance to us. The following
relations that are satisfied by $S^a$ turn out to be useful in the
next section:
\begin{equation}
\sum_{j=1}^{p-q} (-1)^{j+p-q} S^{a}_j=(p^2-q^2),\label{eq:impCoSum}
\end{equation}
\begin{equation}
\sum_{j=p-q+1}^{p} (-1)^{j+p-q} S^{a}_j=q^2. \label{eq:clCoSum}
\end{equation}
The First sum is over (half of) the impurity nodes and the second
sum over (half of) the clean nodes.

\subsection{Adding fractional branes} \label{subsec:ChiAnom}

\noindent In \cite{Klebanov:2002} it was demonstrated that the
chiral anomaly of the cascading gauge theory dual to the KS geometry
\cite{Klebanov:2000} can be understood from the bulk point of view
as Higgs mechanism. The massless scalar in the Betti hyper multiplet
is eaten by the graviphoton (which in the absence of fractional
branes gauges the U(1)$_R$ in the bulk) leading to the bulk vector
acquiring mass and hence the loss of current conservation from the
boundary point of view. It is of no surprise then, that our
identification of the operator dual to the Betti hyper in $Y^{p,q}$
allows us to investigate the effects of adding fractional branes in
these geometries.

One may a priory expect that, similar to what Klebanov and Strassler
did with $T^{1,1}$, one can add fractional branes to $Y^{p,q}$
geometries and study such phenomena as chiral symmetry breaking and
confinement in the related quivers from the gravity side. A
perturbative attempt to construct one such complete supergravity
solution for $Y^{p,q}$ was made in \cite{Burrington:2005zd}, based
on the asymptotic solution of \cite{Herzog:2005}, but their approach
was obstructed by the absence of complex deformations of the
singularity at the tip of the cone over $Y^{p,q}$. This was later
interpreted as absence of a supersymmetric vacuum in such theories
and evidence was proposed for a runaway behavior in the general case
\cite{Berenstein:2005xa,Franco:2005,Bertolini:2005,Brini:2006}.
However, for the $q=0$ cases corresponding to
$T^{1,1}/\mathbb{Z}_p$, confinement and chiral symmetry breaking are
expected for the field theories, and the gravity dual (being a $Z_p$
orbifold of the KS solution) confirms the expectations
\cite{Franco:2005fd}.

In this paper we only use the large-$r$ behavior of the solution
given in \cite{Herzog:2005} as a guide for relating the CFT
couplings and the gravity modulus, similar to what was done in
\cite{Klebanov:2002} (see also
\cite{Bertolini:2001,Bertolini:2002}). We are assuming that this
essentially UV computation is valid despite the out-of-control IR
regime of general cascading $Y^{p,q}$ quivers.

With the aid of our proposal in (\ref{eq:marginalCombination}), and
following \cite{Klebanov:2002} (their equation (18)), we write
\begin{equation}
\sum_{j} S^{a}_j
(\Theta_j-\Theta_{j'})=\frac{1}{\pi\alpha'}\int_{\Sigma_2} C_{2},
\hspace{.7in} \sum_{i} \Theta_{i}\sim C,\label{eq:moduli}
\end{equation}
with $C$ being the RR scalar. In the first equation we have used the
fact that $S^{a}_j=-S^{a}_{j'}$.

We are going to test the relations in (\ref{eq:moduli}) by a
gravitational computation of their RHS and a field theoretical
computation of their LHS. Note that the non-zero value of the $B_2$
field that we referred to in footnote 8 has no effect on
(\ref{eq:moduli}).

\subsubsection{The gravity side}

\noindent The RHS of the second relation in (\ref{eq:moduli}) is
easy to find: the RR scalar is zero, similarly to the case of
$T^{1,1}$ discussed in \cite{Klebanov:2002}.

To find the RHS of the first relation in (\ref{eq:moduli}) we need
$C_2$. Herzog, Ejaz and Klebanov \cite{Herzog:2005} give (in their
equation (41)) the following expression for the RR 3-form in the
background with $M$ fractional branes
\begin{equation*}
F_{3}=\frac{M\alpha'}{2}(p^2-q^2)[\mathrm{d}\psi\wedge\omega_2+\mathrm{d}(\frac{-y\cos\theta}{2(1-y)}\
\mathrm{d}\phi\wedge\mathrm{d}\beta)],
\end{equation*}
with $\omega_2$ a two-form\footnote{$\omega_2$ is related to the
$\omega$ in \cite{Herzog:2005} via $\omega_2=3\omega$. We have also
chosen the opposite sign normalization for $F_3$ so that
$\int_{\Sigma_3} F_3=4\pi^2 \alpha' pM$.} given by
\begin{equation*}
\omega_2=\frac{\sin\theta}{2(1-y)}\mathrm{d}\theta\wedge\mathrm{d}\phi-\frac{1}{2(1-y)^2}\mathrm{d}y\wedge\mathrm{d}\beta-\frac{\cos\theta}{2(1-y)^2}\mathrm{d}y\wedge\mathrm{d}\phi.
\end{equation*}

The 3-form $F_3$ can be locally written as the differential of a
2-form whose \emph{$\psi$-dependent part} is
\begin{equation}
C_{2}=\frac{M\alpha'}{2}(p^2-q^2)\psi\omega_2,\label{eq:C2}
\end{equation}
quite similar to equation (16) in \cite{Klebanov:2002}. Also similar
is the action of the Reeb vector, which is none other than $\psi\to
\psi+2\epsilon$, assuming $\delta \beta=\delta \phi=\delta y=\delta
\theta=0$ \cite{Gauntlett:2004b}.

Equation (\ref{eq:C2}) shows that to evaluate the RHS of the first
relation in (\ref{eq:moduli}) the following integral is needed
\cite{Herzog:2005}
\begin{equation*}
\int_{\Sigma_2} \omega_2=\frac{4\pi
p^2}{3(p^2-q^2)}(p+\sqrt{4p^2-3q^2}).
\end{equation*}
Combining (\ref{eq:C2}), (\ref{eq:moduli}) and the above result for
the integral, we obtain
\begin{equation}
\sum_{j} S^{a}_j (\Theta_j-\Theta_{j'}) = 4M\epsilon p^2
(p+\sqrt{4p^2 -3q^2})/3.\label{eq:gravityResult}
\end{equation}

\subsubsection{The field theory side}

\noindent Now let us do the field theoretical calculation. For
future convenience we define
\begin{equation}
x=\frac{2p-\sqrt{4p^2-3q^2}}{3q^2}.\label{eq:xDefined}
\end{equation}
Note in particular that $x=\frac{1}{4p}$ when $q\rightarrow 0$, and
$x=\frac{1}{3p}$ when $q=p$. The $R$-charges of various
bifundamental fields in the quivers are now expressed as
\begin{equation}
r_U=2px,\quad r_V=1-qx,\quad r_Y=1-(2p-q)x,\quad r_Z=1-(2p+q)x.
\label{eq:biFundamentalRcharges}
\end{equation}

The coefficients $\rho$ appearing in the following are to be
multiplied by
\begin{equation*}
\frac{M}{16\pi^2}(F^{a\ ij}\tilde{F}^{a}_{ij})_{\mbox{node}},
\end{equation*}
and then summed over all the gauge factors to yield the anomalous
divergence of the chiral $R$-current. $\rho$ can also be easily
related to $\Theta$ angles (as in \cite{Klebanov:2002}) via
\begin{equation}
\Theta=2\rho M\epsilon.\label{eq:rhoandTheta}
\end{equation}

A simple field theoretical computation of the chiral anomaly for
general $Y^{p,q}$ (reproduced in appendix \ref{app:Theta}) yields
\begin{equation}
\rho^{p,q}_{\text{\footnotesize{imp }}j}=(-1)^{j+p-q}
\left(p+q^{2}x\right),\label{eq:imYpq}
\end{equation}
and
\begin{equation}
\rho^{p,q}_{\text{\footnotesize{cl }}j}=(-1)^{j+p-q}
\left(p+q^{2}x-2p^{2}x\right).\label{eq:clYpq}
\end{equation}
Note that $(-1)^{j+p-q}$ is positive if the impurity (respectively
clean) node has a bifundamental field $Z$ (respectively $V$)
entering it, and negative otherwise.

Using (\ref{eq:imYpq}) and (\ref{eq:clYpq}), one can obtain from
(\ref{eq:rhoandTheta})
\begin{equation}
\Theta^{p,q}_{\text{\footnotesize{imp}}}=(-1)^{j+p-q}
2M\epsilon\left(p+q^{2}x\right),\label{eq:imYpqTheta}
\end{equation}
and
\begin{equation}
\Theta^{p,q}_{\text{\footnotesize{cl}}}=(-1)^{j+p-q}
2M\epsilon\left(p+q^{2}x-2p^{2}x\right).\label{eq:clYpqTheta}
\end{equation}
%

\begin{figure}[t]
\centering
    \includegraphics[scale=1]{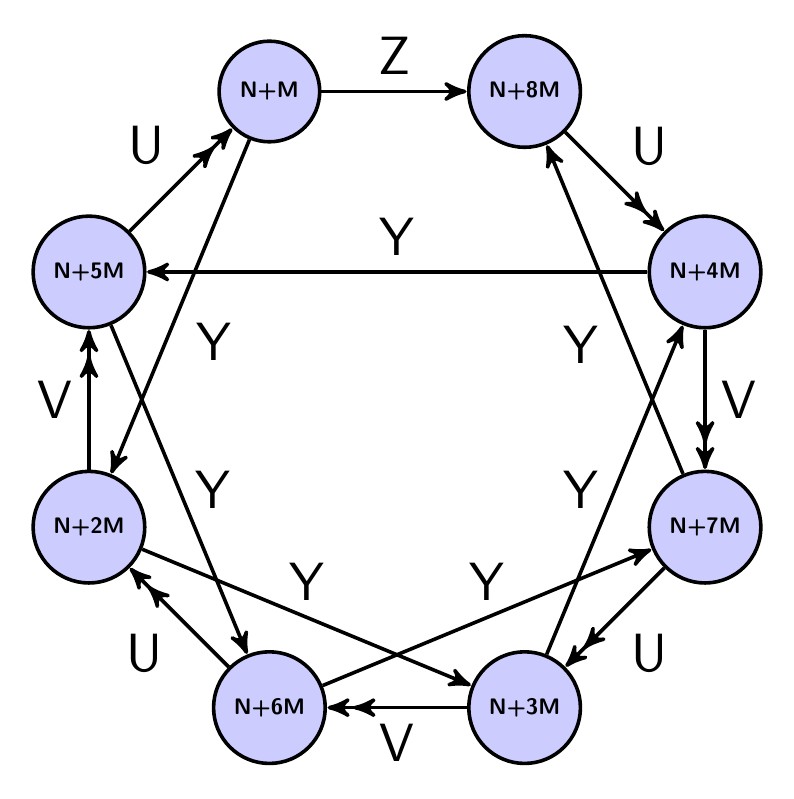}
\caption{The cascading quiver obtained from $Y^{4,3}$.
\label{fig:Y43}}
\end{figure}

As an example, for $Y^{4,3}$ (shown in Figure \ref{fig:Y43}), we
have $x=\sqrt{37}/27$, so
\begin{equation*}
\begin{split}
\partial_{i}J^{i}=&\frac{M}{16\pi^2}[-(p+q^2 x)(F^{a\ ij}\tilde{F}^{a}_{ij})_{N+M}+(p+q^2 x)(F^{a\ ij}\tilde{F}^{a}_{ij})_{N+8M}+\\
&(p+q^2 x-2p^2)(F^{a\ ij}\tilde{F}^{a}_{ij})_{N+ 2M}-(p+q^2
x-2p^2)(F^{a\
ij}\tilde{F}^{a}_{ij})_{N+ 7M}\\
&-(p+q^2 x-2p^2)(F^{a\ ij}\tilde{F}^{a}_{ij})_{N+3M}+(p+q^2
x-2p^2)(F^{a\
ij}\tilde{F}^{a}_{ij})_{N+ 6M}+\\
&(p+q^2 x-2p^2)(F^{a\ ij}\tilde{F}^{a}_{ij})_{N +4M}-(p+q^2
x-2p^2)(F^{a\ ij}\tilde{F}^{a}_{ij})_{N+ 5M}].\\
\end{split}
\end{equation*}

\subsubsection{Consistency of the gravitational and field theoretical
results}

The second relation in (\ref{eq:moduli}) is obviously satisfied
since $\Theta_j$ come in pairs with opposite sign
$\Theta_{j'}=-\Theta_{j}$ and therefore add up to zero, consistent
with vanishing of $C$ in the dual backgrounds.

From (\ref{eq:imYpqTheta}) and (\ref{eq:clYpqTheta}) we can now
check equation (\ref{eq:gravityResult}):
\begin{equation}
\begin{split}
\sum_{j} S^{a}_j (\Theta_j-\Theta_{j'})&= 2(p^2-q^2)
|\Theta_{\mbox{\footnotesize{imp}}}| +2q^2
|\Theta_{\mbox{\footnotesize{cl}}}|\\
&= 4M\epsilon p^2 (p-q^{2}x).\label{eq:matching}
\end{split}
\end{equation}
To write the first equation we have employed the relations
(\ref{eq:impCoSum}) and (\ref{eq:clCoSum}), with the extra fact that
$\Theta_{j'}=-\Theta_{j}$.

Note that if we had not found matching as in (\ref{eq:matching}), we
could have proposed that the missing relative factor must be
inserted in the initial proposals (\ref{eq:BettiDual}) and
(\ref{eq:marginalCombination}). Our success, however, supports the
relations (\ref{eq:BettiDual}) and (\ref{eq:marginalCombination}) as
they are.
%
\section{Further examples with the general algorithm}
\label{sec:algEx}

\noindent In this section we want to explore the difficulties that
arise when searching for exactly marginal operators dual to $B$
field moduli in more general toric geometries than $Y^{p,q}$. A
particularly interesting class of more general toric SE$_5$
manifolds is $L^{a,b,c}$
\cite{Cvetic:2005,Franco:2006,Butti:2005bf}.

Before examining specific examples, let us start by a few general
remarks. One may hope to find exactly marginal directions of general
$L^{a,b,c}$, similar to what we did for $Y^{p,q}$. When smooth,
$L^{a,b,c}$ manifolds have the same topology as $Y^{p,q}$, hence
they possess precisely one B-deformation. We did not succeed in
finding a general expression (something similar to
(\ref{eq:margCos})) for this exactly marginal direction. This is
because we did not manage to find an efficient general
representation for the $L^{a,b,c}$ quivers\footnote{The construction
in \cite{Eager:2010} seems to provide a potentially useful starting
point for finding such a general representation.}. For $Y^{p,q}$
quivers such general representation was explained in Figure
\ref{fig:Y84}. Therefore we now turn to specific members of the
$L^{a,b,c}$ family and look for possibly new features of their
B-deformations.

Our first example in this section is $L^{1,5,2}$. The quiver theory
was given explicitly in the appendix of \cite{Franco:2006}. We start
by forming $B_{K}$ of this quiver. From the general formula
$N_f=a+3b$ \cite{Franco:2006}, we quickly see that $B_{K}$ is a
$16\times 16$ matrix. The null vectors of $B^T_{K}$ give the exactly
marginal directions, as explained in section \ref{sec:NSVZ}. We omit
the details and only report the result. There are three such
vectors. The three dimensional space spanned by these vectors
certainly contains the directions corresponding to the sum of gauge
couplings and the $\beta$-deformation. Therefore, two out of the
three vectors can be safely substituted by $(1,1,...,1)^T$ and
$(0,...,0,1,-1,...,1,-1)^T$, corresponding respectively to the sum
of the gauge couplings and the $\beta$-deformation. A Gram-Schmidt
procedure will then find the combination perpendicular to the
previous two, which is dual to the $B$-field modulus. However,
because the normalization of the null vectors is arbitrary, a
proposal like (\ref{eq:marginalCombination}) can only be made up to
an overall factor. The overall factor can then be determined by
further inspection of the geometries deformed by adding fractional
branes, as in subsection \ref{subsec:ChiAnom}.

As the next example we consider $L^{1,2,1}$, also known as the
Suspended Pinch Point (SPP). The geometry contains a codimension
four singularity and is not smooth \cite{Franco:2006}. Hence, it is
not surprising to see new features arise in this case. The details
of this example are given in the appendix \ref{app:SPP}. In the
following we highlight the procedure.

The related Konishi matrix is $7 \times 7$. After forming $B_{K}$
and finding the null vectors of its transpose, we find four exactly
marginal directions. Two of them are again the sum of the gauge
couplings and the $\beta$-deformation. The other two can be obtained
by a Gram-Schmidt procedure, and are dual to the $B$-field moduli.
The additional one, compared to $L^{1,5,2}$, arises because
$L^{1,2,1}$ is singular and has a fixed circle; the fixed circle
gives rise to a twisted sector that presumably contains (rather
similar to the case of $S^5/\mathbb{Z}_2$ mentioned in subsection
\ref{subsec:stateop}) a Betti hyper multiplet with a modulus inside
it. The remaining piece of work would be to put the two
B-deformations in one-to-one correspondence with the two-cycles in
the dual geometry. This is achieved by comparing
$S^a_{t(I)}-S^a_{h(I)}$ of the B-deformations, with the baryonic
charge assignments $Q_J (X_I)$ of the smooth two-cycle given in
Table 1 of \cite{Franco:2006}. The exactly marginal direction
consistent with the baryonic charge assignment of the smooth
two-cycle corresponds to that two-cycle, and the orthogonal exactly
marginal direction corresponds to the (blown-up) fixed circle.

Similarly, for cases with more than two B-deformations, help from
the geometry side is needed to determine the appropriate baryonic
charge assignments. Then these charge assignments can serve to
disentangle the B-deformations into an orthogonal set whose members
are in a one-to-one correspondence with the non-trivial two-cycles
in the dual geometry.

\section{Summary and discussion}
\label{sec:disc}

\noindent In this paper we simultaneously completed the
identification of the exactly marginal directions of generic
$Y^{p,q}$ theories, and the determination of the protected operators
dual to their shortened supergravity multiplets. The exactly
marginal operator that we have found is dual to the $B$-field
modulus of the gravity side. This modulus is incarnated at the
linear level as a scalar component of a Betti hyper multiplet in the
supergravity KK spectrum. The superfield version of the exactly
marginal operator is thus dual to the Betti hyper multiplet.

We found the exactly marginal direction from the NSVZ equations, in
the language developed in \cite{Imamura:2007}. In this approach,
which applies to gauge theories described by brane tiling, finding
exactly marginal operators boils down to solving a system of
difference equations with coefficients $\pm 1$ (or $2$ if there are
adjoint chiral fields in the gauge theory). We showed that the
solutions to these equations can be thought of as left null vectors
of a neatly derivable matrix, that we referred to as the Konishi
matrix of the quiver, and denoted by $B_{K}$. The left nullity of
the Konishi matrix (which equals its nullity, since the matrix is
$N_f\times N_f$) is $2+b_3($SE$_5)$: one null direction corresponds
to the sum of the gauge couplings, another corresponds to the
$\beta$-deformation, and the remaining $b_3($SE$_5)$ correspond to
the exactly marginal deformations dual to the $B$-field moduli. We
called the last set \emph{B-deformations}. Unlike the
$\beta$-deformation, B-deformations do not break any global
symmetry. We saw in section \ref{sec:Ypq} that B-deformations
\emph{generically} involve tuning the superpotential couplings. It
is not difficult to show that they \emph{always} involve tuning of
at least some of the gauge couplings; this is proved in appendix
\ref{app:Konishiproofs}.

We further pointed out that any set of baryonic charge assignments
to matter fields gives a right null vector of $B_{K}$, but not every
right null vector of $B_{K}$ yields a consistent baryonic charge
assignment. This is because $B_{K}$ encodes only local data on the
tiling. The two non-trivial cycles of the torus on which the tiling
is defined impose two additional consistency relations on the
baryonic charge assignments. Thus, only a codimension two subspace
of the null space of $B_{K}$ corresponds to consistent baryonic
charge assignments. This conclusion is obvious in retrospect as a
codimension two subspace of the null space of $B_{K}$ would be
$b_3($SE$_5)$ dimensional, and this is the number of global baryonic
U(1) symmetries of the field theory.

In appendix \ref{app:Konishiproofs} it is shown that the Konishi
matrix can alternatively be thought to arise from Konishi anomaly
equations. This point of view is advantageous, as compared to that
of the NSVZ equations, in that it helps to frame the analysis in the
context of the chiral ring of the field theory. Also, this viewpoint
enables us to recognize the usefulness of the Konishi matrix beyond
toric gauge theories.

There are various problems that follow naturally from our
investigation. One important issue which deserves further study is
that relation (\ref{eq:marginalCombination}) is only correct up to
an additive constant that we have not been able to compute; see our
comment in footnote 8. It would be nice to have a systematic
approach to compute this constant for arbitrary toric theories.
Another problem is that we have not found a solid argument to
support our conjecture, presented in (\ref{eq:primaryMarginal1}),
for the form of the exactly marginal primary operators that deform
gauge and superpotential couplings. We have provided partial support
for our conjecture below equation (\ref{eq:Y84BettiDual}), and also
in appendix \ref{app:Konishiproofs}. However, it would be highly
desirable to have a sharp argument establishing (or ruling out) the
form (\ref{eq:primaryMarginal1}) for these operators.

In this paper we focused only on exactly marginal deformations that
can be obtained by changing couplings already present in the
original theories. As mentioned at the end of section \ref{sec:NSVZ}
and in subsection \ref{subsec:stateop}, more exactly marginal
directions may exist, which following \cite{Imamura:2007} we
referred to as ``accidentally marginal''. These would arise from
mesonic exactly marginal chiral primary operators absent in the
superpotential. For example, the conifold theory, the
$\mathcal{N}=4$ theory, and $Y^{2,1}$ quiver theory, with respective
global non-R symmetries SU(2)$\times$SU(2), SU(3), and
SU(2)$\times$U(1), have respectively two, one, and zero accidentally
marginal directions. It would be interesting to study theories
admitting such accidentally marginal deformations to see if the
following rule of thumb can be made more precise: \emph{a larger
global symmetry group yields a larger conformal manifold}. A precise
version of the previous statement was conjectured by Kol
\cite{Kol:2002}, in a form that neglects B-deformations and the
exactly marginal direction dual to the axion-dilaton. For the
remaining directions (including the $\beta$-deformation)
\cite{Kol:2002} realizes the importance of the symmetric
representation of the global flavor group in determining the
dimensionality of the conformal manifold. However, as it stands, the
conjecture of \cite{Kol:2002} is not correct for the known toric
quivers, and it is not clear how to amend it. Therefore, it seems
that more work is required to make the above rule of thumb
precise\footnote{We hasten to add that it seems the dimension of the
symmetric representation of the global non-$R$ symmetry group of the
field theory (that we shall denote by dim(sym$(F)$)) can give a
quantity with which to consistently (with the known examples) define
the word \emph{larger} in the rule of thumb: \emph{larger global
symmetry group} can be taken to mean greater
dim(sym$(F))+$rank$(F)$; dim(sym$(F)$) might be related to the
number of accidental exactly marginal deformations, and rank$(F)$
gives the number of non-accidental ones in a toric quiver theory.}.
To that end, the analysis of \cite{Green:2010} would arguably play a
key role, but needs to be supplemented with a method to first obtain
the number of marginal chiral primary operators of a quiver. Note
that since according to \cite{Green:2010} the global flavor group
can make some of the marginal chiral primaries irrelevant (in a
manner rather analogous to the Higgs mechanism), our rule of thumb
needs that the number of marginal chiral primaries grow fast enough
with the size of the flavor group to (over)compensate the loss of
exactly marginal primaries; although this is the case in all the
examples we are aware of, we have no proof why this should be true
in general. We hope to report more progress on this in the future.

\acknowledgments

\noindent We thank R.~Eager for correspondence, C.~Herzog for
sharing with us unpublished notes on cascading $Y^{p,q}$ quivers,
and Y.~Tachikawa for an in-depth discussion on the chiral anomaly of
cascading $Y^{p,q}$ quivers back in 2007. We also thank
M.~Bertolini, S.~J.~Rey, and Y.~Tachikawa for their insightful
comments on an earlier version of this paper. A.~A.~A is also
grateful to A.~Hanany and J.~T.~Liu for their helpful suggestions
for this project. This work was supported in part by DoE grant
DE-FG02-95ER40899.

\appendix

\section{Proofs for the properties of the Konishi matrix} \label{app:Konishiproofs}

In section \ref{sec:NSVZ} it was stated that every baryonic charge
assignment satisfying (\ref{eq:bChargeCond}) gives an $N_f$ tuple
$(Q_J(X_1),...,Q_J(X_{N_f}))^T$ that is a null vector of $B_{K}$.
This is seen to be true by noting that any of the first $N_g$ rows
of
\begin{equation}
B_{K}\times (Q_J(X_1),...,Q_J(X_{N_f}))^T=0 \label{eq:BQ}
\end{equation}
imposes the first condition in (\ref{eq:bChargeCond}) for the
corresponding node in the quiver, while any of the last $N_W$ rows
of (\ref{eq:BQ}) imposes the second condition in
(\ref{eq:bChargeCond}) for the corresponding superpotential loop in
the quiver.

The second property of $B_{K}$ cited in the main text is that every
set of coefficients $S^a$, $S^k$ that gives an exactly marginal
direction of the form (\ref{eq:ImamuraConventionRep}), yields an
$N_f$ tuple
\begin{equation*}
(S^a_1 ,...,S^a_{N_g},S^k_1,...,S^k_{N_W})
\end{equation*}
that is a left null vector of $B_{K}$. This is true because every
column of
\begin{equation}
(S^a_1 ,...,S^a_{N_g},S^k_1,...,S^k_{N_W})\times B_{K}=0
\label{eq:SB}
\end{equation}
is equivalent to the relation (\ref{eq:ImamuraRelation}) for the
corresponding chiral field (recall that columns of $B_{K}$ are
labeled by the chiral fields in the quiver).

We now show\footnote{We thank Y.~Tachikawa for pointing out the
following neat Konishi anomaly argument to us.} the important fact
that the left null vectors of $B_{K}$ are in one-to-one
correspondence with the chiral primaries that one can form with
Tr$W_j^2$ and $W_m$. To see this, consider a chiral bifundamental
field $\Phi_I$ (the modifications required for adjoint chiral fields
are straightforward), and define $J_I=\mathrm{Tr}\Phi_I
e^{V_{h(I)}}\bar{\Phi}_I e^{-V_{t(I)}}$. Then from the Konishi
anomaly equation we have
\begin{equation}
\frac{32\pi^2}{N}\bar{D}^2 J_I = \sum_{j\in I} (Tr W_j^2)
+\sum_{m\in I} (\frac{32\pi^2}{N}W_m). \label{eq:Konishi}
\end{equation}
Notice that the coefficients on the RHS are the entries of the
Konishi matrix in the $I$th column, except for the reversed sign of
the $W_m$ coefficients. A linear combination of Tr$W_j^2$ and $W_m$
that is a chiral primary should be perpendicular to the RHS of
(\ref{eq:Konishi}) for every $I$. There are $N_f$ relations like
(\ref{eq:Konishi})---one for each $I$. There are also $N_f$
operators of the form Tr$W_j^2$ or $W_m$---$N_g$ of the former, and
$N_W$ of the latter. Thus if the RHS of (\ref{eq:Konishi}) for every
$I$ gave a different expression, the orthogonalization procedure
would leave no linear combination of Tr$W_j^2$ and $W_m$ as a chiral
primary. But every time a linear combination of $\bar{D}^2 J_I$
vanishes, there is one fewer constraint on the chiral primary
combinations of Tr$W_j^2$ and $W_m$, and therefore one more of such
chiral primaries. That these marginal chiral primary operators are
indeed exactly marginal can then be deduced from either NSVZ,
AdS/CFT, or a symmetry analysis as in \cite{Green:2010}. Note that
if we knew how to perform the orthogonalization procedure mentioned
above, it would give us the correction terms proportional to $W_m$
in the chiral primary combinations, and that would yield the
required tuning of the superpotential couplings on the conformal
manifold. But at least for a B-deformation with $S^k_m=0$ (as in
that of $Y^{p,0}$), it already seems natural to expect that
operators of the form (\ref{eq:primaryMarginal1}) are perpendicular
to all $\bar{D}^2 J_I$ in (\ref{eq:Konishi}).

Despite our inability to carry out the orthogonalization procedure
referred to earlier, we conjecture that the primary operators
perpendicular to (\ref{eq:Konishi}) are of the form
\begin{equation}
\sum_{j} S^a_j(Tr W_j^2) -\sum_{m} S^k_m(\frac{32\pi^2}{N}W_m).
\label{eq:primaryForm}
\end{equation}
This conjecture is motivated by the following argument. At weak
coupling, $g'$ (defined below equation (\ref{eq:ImamuraConvention}))
can be identified with the holomorphic coupling $g_h$. Then it is
not difficult to see that small variations in one of the
combinations (\ref{eq:ImamuraConvention}), while keeping constant
the rest, are indeed generated by adding operators of the form
(\ref{eq:primaryForm}) to the superpotential. The apparent mismatch
between the factor of 8 in (\ref{eq:ImamuraConvention}) and the
factor of 32 in (\ref{eq:primaryForm}) is explained by noting that
$\mathcal{L}_h = \frac{1}{4} \int \mathrm{d}^2\theta W^2/g_h^2$.
Note also that (\ref{eq:primaryForm}) gives the expected operator
for the $\beta$-deformation, which has $S^a_j=0$ and $S^k_m=\pm1$.
This provides the further partial support for
(\ref{eq:primaryMarginal1}) that we promised in the main text.

Now, since $B_{K}$ is a square matrix, its left and right nullities
are equal. This, however, does not mean that there are as many
exactly marginal directions in the quiver as there are consistent
baryonic charge assignments. The reason is that \emph{not every
right null vector of $B_{K}$ gives a consistent baryonic charge
assignment}. This is because equation (\ref{eq:BQ}) imposes the
second condition in (\ref{eq:bChargeCond}) only on the
superpotential loops in the quiver. As we show below, to find
consistent baryonic charge assignments one should supplement
(\ref{eq:BQ}) with two more relations, and hence only a codimension
two subspace of the (right) null space of $B_{K}$ corresponds to
consistent baryonic charge assignments.

One way to understand why the number two comes in, is to realize
that ensuring the second condition in (\ref{eq:bChargeCond}) on all
loops in the quiver requires supplementing (\ref{eq:BQ}) by two
relations arising from the two non-trivial cycles in the torus of
brane tiling. To see this, note that any of the last $N_W$ rows of
(\ref{eq:BQ}) imposes the second condition in (\ref{eq:bChargeCond})
for the corresponding superpotential node in the tiling. Since brane
tilings define bipartite graphs, each edge (i.e. chiral field) can
be assigned a direction (e.g. from black nodes to white nodes)
\cite{Franco:2006bd}. With such directions assigned to the edges in
brane tiling, one can interpret the second condition in
(\ref{eq:bChargeCond}) as Kirchhoff's current law for arbitrary
Gauss surfaces (that correspond to arbitrary loops in the quiver
diagram) in `the brane tiling circuit'. We are thus interpreting the
baryonic charge of a chiral field as the current its corresponding
edge carries on brane tiling\footnote{Incidentally, $S^a$
coefficients can be interpreted as the currents circulating in the
loops of brane tiling. This follows from the equation
$S^a_{t(I)}-S^a_{h(I)}=Q_J(X_I)$ \cite{Imamura:2007}, mentioned in
section \ref{sec:NSVZ}, that relates the B-deformations to their
corresponding baryonic charge assignments.}. Now, equation
(\ref{eq:BQ}) ensures Kirchhoff's current law on every node in the
tiling, because nodes correspond to superpotential loops in the
quiver. It is clear that this guarantees Kirchhoff's current law for
all shrinkable Gauss surfaces on the tiling. But to ensure the full
consistency of the corresponding baryonic charge assignment, one has
to add the two Kirchhoff current laws arising from the two
non-trivial cycles in the torus on which the tiling is
defined\footnote{In the circuit language, this means that no net
current should be carried along the periodic directions.}. These are
the two relation that supplement (\ref{eq:BQ}) to give fully
consistent baryonic charge assignments.

Instead of the argument of the previous paragraph, one could again
use (\ref{eq:Konishi}) to verify that for every null vector of
$B_{K}$ there is one conserved current in the form of a linear
combination of $J_I$, but two of the conserved currents are those of
the global U(1)$\times$U(1) flavor symmetry of the CFT (equation
(\ref{eq:U1current}) gives an example). Therefore a two dimensional
subspace of the null space of $B_{K}$ corresponds to flavor U(1)
charge assignments, and the rest of it to the baryonic charge
assignments. The relation between the flavor U(1) symmetries and the
non-trivial cycles of the tiling (which played a key role in the
argument of the previous paragraph) is well-known (see for example
\cite{Kennaway:2007}). Also, from this argument it becomes clear in
what sense the global non-$R$ U(1) symmetries are responsible for
the $2+b_3($SE$_5)$ exactly marginal directions of toric quivers.

Finally, we show that B-deformations always involve tuning gauge
couplings. In other words, there are no B-deformations with all
their $S^a$ coefficients equal to zero. Let us assume there is one
such deformation. Then starting with a node $P$ on brane tiling and
considering the relation (\ref{eq:ImamuraRelation}) for an edge $I$
connected to it, we see that the $S^k$ coefficient of the node $Q$
at the other end of $I$ should be negative of the $S^k$ coefficient
of $P$. Then considering (\ref{eq:ImamuraRelation}) for another edge
$I'$ connected to $Q$ and so on, we see that the $S^k$ coefficients
on the tiling only alternate signs. This means we end up with the
$\beta$-deformation (up to an insignificant normalization which is
the value of $S^k$ chosen for the initial node $P$). Hence this is
not a B-deformation.

To summarize, the Konishi matrix encodes local data on brane tiling.
This local data is sufficient (and necessary) for determining the
exactly marginal directions that we are concerned about (recall that
in the present paper we are not concerned about the accidentally
marginal directions, referred to at the end of section
\ref{sec:NSVZ}); these exactly marginal directions can be obtained
from left null vectors of $B_{K}$. However, to determine the
consistent baryonic charge assignments, the local data in $B_{K}$
(although necessary) must be supplemented by the global data encoded
in the two nontrivial cycles of the torus of brane tiling; thus
baryonic charge assignments form a codimension two subspace of the
right null space of $B_{K}$.

\section{Field theoretical computation of $\Theta ^{p,q}$ for the cascading $Y^{p,q}$ quivers}
\label{app:Theta}

In this appendix, we prove the field theoretical relation
(\ref{eq:imYpq}). Equation (\ref{eq:clYpq}) is derived similarly.

\begin{figure}[t]
\centering
    \includegraphics[scale=1]{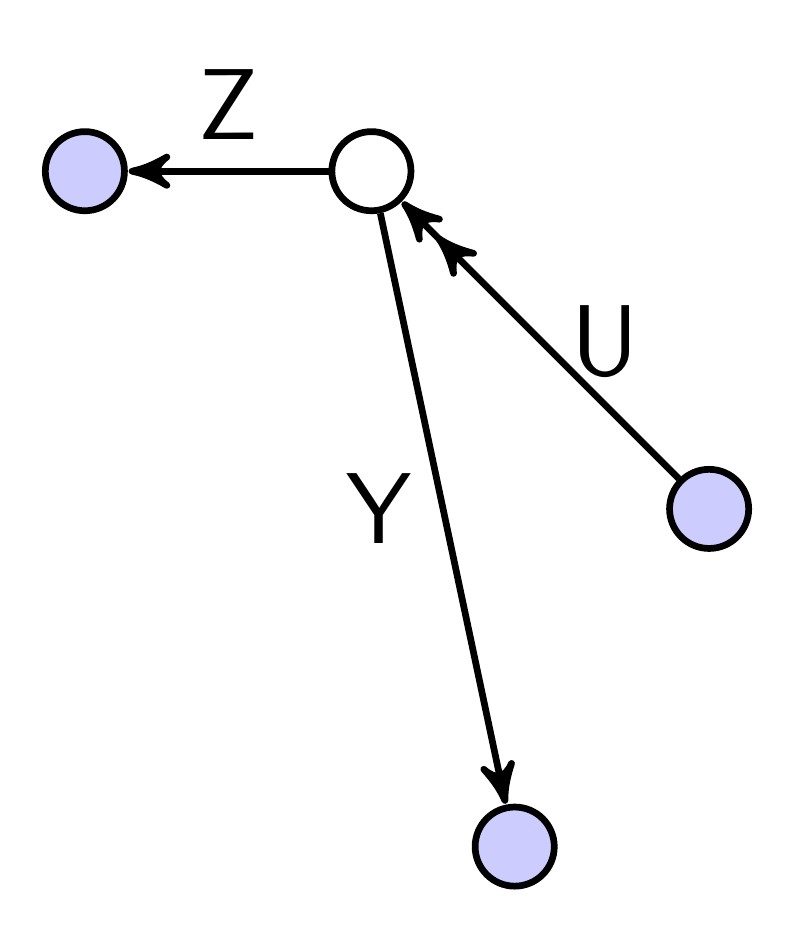}
\caption{A typical impurity node in $Y^{p,q}$ quivers is shown as an
empty node, with all its neighbors attached to it via chiral
bifundamentals. \label{fig:impapp}}
\end{figure}

Take an impurity node $P$ that has a gauge factor of rank $N+k M$
with some $k$; this could be the empty node in Figure
\ref{fig:impapp}. Assume that the node has a bifundamental field
singlet $Z$ `exiting' it. This bifundamental would enter a node with
rank $N+kM+pM+qM$, as dictated by the baryonic charge of $Z$
\cite{Benvenuti:2005b}. There is also a bifundamental doublet $U$
entering $P$, which emanates from a node with rank $N+kM+pM$ as
dictated by the baryonic charge of $U$. Finally, a bifundamental
singlet $Y$ leaves $P$ to a destination node with rank $N+kM+pM-qM$
as dictated by the baryonic charge of $Y$. The chiral anomaly
($\mathrm{Tr}R$) of the fermions charged under $P$ is then given by
\begin{equation*}
\begin{split}
&\frac{1}{2}\times(N+kM+pM+qM)(r_Z-1)+2\times\frac{1}{2}\times(N+kM+pM)(r_U-1)\\
&+\frac{1}{2}\times(N+kM+pM-qM)(r_Y-1)+N+kM\\
&\ =-(p+q^2 x)M,
\end{split}
\end{equation*}
where we have used (\ref{eq:biFundamentalRcharges}). In the above
equation, the factors $\frac{1}{2}$ for the first three terms on the
LHS are the Dynkin index of the fundamental representation, the
extra coefficient $2$ for the second term is because $U$ is a
doublet, and the fourth term is the gaugino contribution. A similar
computation for a node which has a bifundamental field singlet $Z$
entering it yields the opposite sign, hence proving
(\ref{eq:imYpq}). The factors $(-1)^{j+p-q}$ appear because of the
way we have numbered the nodes (see the caption of Figure
\ref{fig:Y84}).

\section{Exactly marginal directions for SPP} \label{app:SPP}

In this appendix we form the Konishi matrix of SPP quiver and obtain
from it the exactly marginal directions in the space of couplings.

\begin{figure}[t]
\centering
    \includegraphics[scale=1.2]{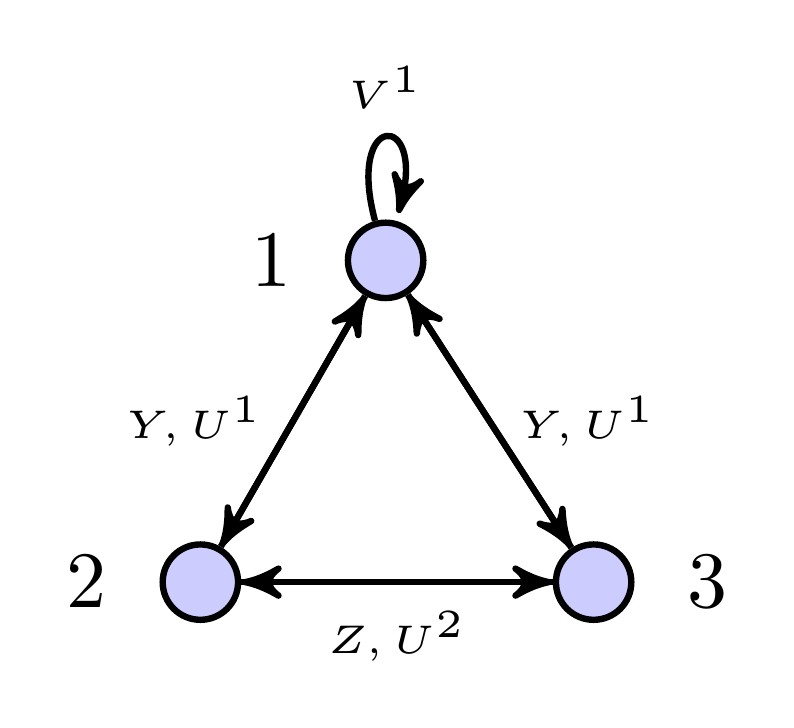}
\caption{The quiver diagram for SPP. The chiral fields are named
following \cite{Franco:2006}. \label{fig:SPP}}
\end{figure}

The quiver is shown in Figure \ref{fig:SPP}. It contains seven
chiral fields
\begin{equation*}
V^1 _{11}, Y_{12}, U^1 _{13}, U^1 _{21}, U^2 _{23}, Y_{31}, Z_{32}.
\end{equation*}

The superpotential terms are \cite{Franco:2006}
\begin{equation*}
\begin{split}
&W_1 =U^1_{21}Y_{12}U^2_{23}Z_{32},\quad
W_2=-Z_{32}U^2_{23}Y_{31}U^1_{13},\\
&W_3=U^1_{13}Y_{31}V^1_{11},\quad W_4=-Y_{12}U^1_{21}V^1_{11}.
\end{split}
\end{equation*}

The Konishi matrix then follows to be
\begin{equation}
B_{K}=
\begin{pmatrix}
  2 & 1 & 1 & 1 & 0 & 1 & 0 \\
  0 & 1 & 0 & 1 & 1 & 0 & 1 \\
  0 & 0 & 1 & 0 & 1 & 1 & 1 \\
  0 & -1 & 0 & -1 & -1 & 0 & -1 \\
  0 & 0 & -1 & 0 & -1 & -1 & -1 \\
  -1 & 0 & -1 & 0 & 0 & -1 & 0 \\
  -1 & -1 & 0 & -1 & 0 & 0 & 0
 \end{pmatrix}.
\end{equation}

The four left null vectors are
\begin{eqnarray}
(S^a _1,S^a _2,S^a _3,S^k_1,S^k_2,S^k_3,S^k_4)=&&(1,1,1,1,1,1,1),\nn\\
&&(0,0,0,1,-1,1,-1),\nn\\
&&(0,2,-2,1,-1,-1,1),\nn\\
&&(4,-3,-3,-3,-3,4,4).\nn
\end{eqnarray}

The first two clearly correspond to the sum of gauge couplings and
the $\beta$-deformation. The last two are B-deformations. The third
one is consistent with the baryonic charge assignments for the
smooth two-cycle (given in Table 1 of \cite{Franco:2006}) and is
therefore dual to the vev of the complex $B$ field on the smooth
two-cycle. The last one is dual to the modulus inside the Betti
hyper multiplet in the twisted sector arising from the fixed circle
of $L^{1,2,1}$.


\end{document}